\documentclass[aps, prb, reprint, superscriptaddress, longbibliography, floatfix]{revtex4-2}
\usepackage{graphicx}% Include figure files
\newcommand{\safeincludegraphics}[2][]{%
  \IfFileExists{#2}{%
    \includegraphics[#1]{#2}%
  }{%
    \fbox{\parbox[c][0.16\textheight][c]{0.90\linewidth}{%
      \centering Missing figure file:\\[0.5ex]\texttt{\detokenize{#2}}}}%
  }%
}
\usepackage{dcolumn}% Align table columns on decimal point
\usepackage{bm}% bold math
\usepackage{physics}
\DeclareMathAlphabet{\pazocal}{OMS}{zplm}{m}{n}
\usepackage{amsmath}
\usepackage{amsfonts}
\usepackage{amssymb}
\usepackage{mathrsfs}
\usepackage{subfigure}
\usepackage{color}
\usepackage{xcolor}
\usepackage{tikz}
\usepackage[colorlinks=true,linkcolor=blue,anchorcolor=red,citecolor=blue, urlcolor=blue]{hyperref}

\begin{document}
\preprint{APS/123-QED}
 
\title{Cooperative Domain-Wall Dynamics in a Two-Dimensional Quasiclassical Holstein Model}
%\title{Charge-Constrained Coarsening in a Two-Dimensional Quasiclassical Holstein Model}

\author{Arunangshu Bora}
\affiliation{Department of Physics, University of Virginia, Charlottesville, Virginia, 22904, USA}

\author{Sankha Subhra Bakshi}
\affiliation{Department of Physics, University of Virginia, Charlottesville, Virginia, 22904, USA}

\author {Ho Jang}
\affiliation{Department of Physics, University of Virginia, Charlottesville, Virginia, 22904, USA}

\author {Gia-Wei Chern}
\affiliation{Department of Physics, University of Virginia, Charlottesville, Virginia, 22904, USA}

\begin{abstract}
We investigate charge-density-wave (CDW) coarsening in a two-dimensional quasiclassical Holstein model following thermal quenches into the ordered phase. Although the staggered CDW order parameter is nonconserved, domain-wall motion remains constrained by conservation of the microscopic electronic charge. The simulations exhibit apparent growth exponents between $1/4$ and the Allen--Cahn value $1/2$. These apparent exponents are primarily accounted for by the crossover kinetics $\dot{L}=[D(T)/L][1+L_\times^2(T)/L^2]$, where $L_\times(T)$ is a temperature-dependent crossover length. At low temperature, nearly binary occupations require compatible wall segments to rearrange cooperatively, producing the $L(t)\sim t^{1/4}$ limit. Thermal broadening creates partially occupied interfacial sites that permit local curvature-driven motion and restore $L(t)\sim t^{1/2}$ at late times. Fits to the measured growth rate, a dimensionless rate collapse, and a long-time large-system simulation consistently support this crossover picture. These results show that a microscopic conservation law can generate long-lived anomalous coarsening even when the emergent order parameter is nonconserved.
\end{abstract}

\date{\today}
\maketitle

\section{Introduction}
\label{sec:introduction}

Phase ordering following a rapid quench is a paradigmatic nonequilibrium process in which locally ordered regions form and subsequently grow through the motion and annihilation of defects~\cite{Bray1994,Onuki2002,SanjayPuri2009,Cugliandolo2015}. At late times, the dynamics is often characterized by a single growing length scale,
\begin{equation}
L(t)\sim t^{1/z},
\end{equation}
representing the typical domain size~\cite{Majumdar1995,Sire1995}. The dynamical exponent $z$ is determined largely by the symmetry and conservation properties of the order parameter~\cite{Hohenberg1977,Bray1994}. For a scalar nonconserved order parameter in two or more dimensions, interfacial curvature drives domain-wall motion, leading to the Allen--Cahn growth law $L(t)\sim t^{1/2}$, or $z=2$~\cite{Allen1979,Lifshitz1962,Bray1994}. When the order parameter is conserved, domain growth instead requires transport over increasingly large distances and is generally slower, with the characteristic law $L(t)\sim t^{1/3}$, corresponding to $z=3$~\cite{Lifshitz1961,Wagner1961,Cornell1991,Bray1994}. These results form the standard framework for classifying phase-ordering kinetics.

In many electronically ordered systems, the slow collective degrees of freedom are coupled to microscopic electronic quasiparticles whose redistribution accompanies the evolution of the order parameter~\cite{Gruner1988,Gruner1994}. This situation arises naturally in electron--lattice models such as the Holstein model~\cite{Holstein1959a,Holstein1959b,Esterlis2019} and Jahn--Teller models~\cite{Millis1995,Millis1996a,Millis1996b,Roder1996,Popovic2000,Hotta2000}, as well as in itinerant magnetic systems such as $s$--$d$ or Kondo-lattice models~\cite{Zener1951,Martin2008,Akagi2010,Chern2010,Azhar2017}, where classical lattice or spin variables evolve under forces mediated by an electronic subsystem~\cite{Ghosh2024,Zhang2023,Cheng2023a}. Even when the symmetry and conservation law of the coarse-grained order parameter suggest a conventional phase-ordering universality class, the microscopic electronic dynamics can introduce additional kinetic constraints, nonlocal couplings, or configuration-dependent mobilities~\cite{Zhang2021,Zhang2022b}. Understanding how such effects modify domain growth is therefore an important problem in the nonequilibrium dynamics of electronically ordered systems~\cite{Ghosh2024,Fan2024a}.

The Holstein model provides a particularly simple setting in which this interplay can be examined. At half filling on a bipartite lattice, the low-temperature phase is a charge-density wave (CDW) with a discrete Ising-like order parameter~\cite{Scalettar1989,Noack1991,Zhang2019,Chen2019,Hohenadler2019}. Since the staggered CDW order is nonconserved, conventional phase-ordering theory would suggest Allen--Cahn growth in two dimensions~\cite{Allen1979,Lifshitz1962,Bray1994}. However, the local CDW order is built from the electronic density, while the total electron number remains conserved. Motion of a domain wall therefore requires a redistribution of charge in the underlying electronic subsystem, providing a microscopic mechanism by which the electronic degrees of freedom can constrain the otherwise nonconserved order-parameter dynamics.

Previous large-scale simulations have already revealed signatures of such nonstandard kinetics in Holstein-type models~\cite{Cheng2023b,Yang2024a}. Machine-learning-enabled simulations of the two-dimensional semiclassical Holstein model found substantial deviations from the conventional Allen--Cahn growth law during CDW coarsening~\cite{Cheng2023b,Dinh2024}. In our subsequent study of the one-dimensional Holstein model, both the full semiclassical model and its quasiclassical strong-coupling limit exhibited anomalously slow domain growth with an apparent temperature-dependent dynamical exponent~\cite{Jang2026}. In one dimension, CDW domains are separated by particle-like kinks, and conventional coarsening proceeds through kink diffusion and pair annihilation~\cite{Glauber1963,Torney1983,Amar1990,Derrida1995,Derrida1996,Krapivsky2010b}. The anomalous behavior was traced instead to cooperative kink motion imposed by electron-number conservation: displacement of an isolated kink changes the electronic occupation and must be accompanied by a compensating rearrangement elsewhere~\cite{Jang2026}. Consequently, the effective kink diffusivity depends on the kink density and decreases as the defects become more dilute~\cite{Jang2026,Krapivsky2012}.

How this cooperative mechanism manifests itself in two dimensions is considerably less obvious. The elementary defects are now extended domain walls rather than isolated kinks, and conventional coarsening is driven by local curvature~\cite{Allen1979,Lifshitz1962,Bray1994}. Moreover, a two-dimensional wall contains a variety of local environments---straight segments, corners, protrusions, and narrow structures---which provide distinct possibilities for rearranging charge while moving the interface. Spatially separated portions of the same wall, or segments belonging to different walls, may also participate in a cooperative update. The availability of such compatible partners therefore changes continuously as the domain pattern coarsens. The central question is thus not simply whether the one-dimensional anomalous exponent survives in two dimensions, but how microscopic charge conservation modifies the curvature-driven motion of extended domain walls.

To address this problem, we consider the two-dimensional Holstein model in the quasiclassical limit introduced in our previous work~\cite{Jang2026}. Starting from the semiclassical Holstein model with classical lattice displacements, this limit arises in the strong-coupling regime where the electronic hopping is small compared with the local electron--lattice energy scale, while the electronic subsystem remains fast relative to the lattice dynamics. The resulting description retains the fermionic occupation constraint and the coupling between local charge and lattice distortion, but replaces the itinerant electronic problem by local quasi-equilibrium occupations. This reduction makes large-scale simulations of domain growth feasible while preserving the microscopic charge-conservation effects that were found to control the anomalous kink dynamics in one dimension~\cite{Jang2026}. Here we use this quasiclassical limit to investigate how the same constraint modifies the curvature-driven motion of extended domain walls in two dimensions.

Our simulations reveal that the domain growth is not governed by a single anomalous power law. Instead, the coarsening rate contains both the conventional curvature-driven contribution underlying the Allen--Cahn law and an additional contribution associated with charge-conserving cooperative wall motion. The former gives the familiar $L(t)\sim t^{1/2}$ behavior, whereas the cooperative process requires compatible pairs of local wall rearrangements and produces a stronger scale dependence corresponding to $L(t)\sim t^{1/4}$. The observed temperature-dependent effective exponents therefore arise from a crossover between these two contributions rather than from distinct asymptotic power-law regimes. At low temperature, electronic occupations are nearly binary and charge conservation strongly restricts local wall motion, making cooperative updates important. As the temperature is raised, partially occupied sites become increasingly common along the domain walls, providing local channels for charge redistribution and progressively restoring conventional single-wall updates and Allen--Cahn coarsening.

The remainder of the paper is organized as follows. Section~\ref{sec:qcl} derives the quasiclassical limit of the Holstein model used in our simulations. Section~\ref{sec:coarsening} describes the thermal-quench protocol and presents the numerical results for CDW domain growth. In Sec.~\ref{sec:coarsening_mechanism}, we develop the cooperative domain-wall mechanism and the resulting crossover description. Section~\ref{sec:conclusion} summarizes our conclusions and discusses future directions. The dimensionless formulation used in the simulations is detailed in Appendix~\ref{app:units}.

\section{quasiclassical limit of the Holstein model}
\label{sec:qcl}

We begin with the semiclassical spinless Holstein model on a two-dimensional square lattice,
\begin{align}
\mathcal{H}_{\rm Hol}
={}&
-t_{\rm nn}\sum_{\langle ij\rangle}
\left(c_i^\dagger c_j+c_j^\dagger c_i\right)
-g\sum_i\left(n_i-\frac{1}{2}\right)Q_i
\nonumber\\
&+
\sum_i\left(\frac{P_i^2}{2M}+\frac{K}{2}Q_i^2\right)
+\kappa\sum_{\langle ij\rangle}Q_iQ_j .
\label{eq:holstein}
\end{align}
Here $c_i^\dagger$ and $c_i$ are fermionic creation and annihilation operators, $n_i=c_i^\dagger c_i$, and $t_{\rm nn}$ is the nearest-neighbor hopping amplitude. The variables $Q_i$ and $P_i$ describe a local lattice distortion and its conjugate momentum, with ionic mass $M$ and elastic constant $K$. The last term represents an intersite elastic coupling; for $\kappa>0$, neighboring distortions favor opposite signs.

Treating $Q_i$ as a classical variable is appropriate in the adiabatic regime where the relevant lattice motion is slow compared with the electronic dynamics. Physically, $Q_i$ may represent the amplitude of a local symmetry-preserving structural mode, for example an $A_{1g}$-like breathing distortion of an $MO_6$ octahedron. More generally, the semiclassical Holstein model provides an effective description of slowly evolving local lattice coordinates coupled to a fast electronic subsystem. Such classical-lattice formulations have been widely used for studying structural and charge-ordering dynamics in electron--phonon systems~\cite{Holstein1959a,Holstein1959b,Jang2026}.

At half filling, the square-lattice Holstein model develops a checkerboard charge-density wave (CDW). In the weak-coupling picture, the instability originates from perfect nesting of the half-filled nearest-neighbor band at wave vector $(\pi,\pi)$, which enhances the electronic susceptibility and favors a lattice distortion with the same ordering wave vector. The corresponding charge modulation may be written as
\begin{equation}
\langle n_i\rangle
=
\frac{1}{2}
+
(-1)^{x_i+y_i}\phi ,
\label{eq:cdw_density}
\end{equation}
where the two signs of $\phi$ describe the two symmetry-related checkerboard states.

The same ordered pattern emerges from the strong-coupling viewpoint. A sufficiently large local lattice distortion produces a polaronic energy scale $E_{\rm p}\sim g^2/K$, while virtual hopping generates an effective intersite coupling of order $J_{\rm eff}\sim t_{\rm nn}^2/E_{\rm p}$. At half filling, this favors alternating charge-rich and charge-poor sites accompanied by staggered lattice distortions. Thus the weak- and strong-coupling descriptions lead to the same Ising-like checkerboard CDW, although the microscopic origin is described differently in the two limits.

We consider the adiabatic dynamics of the lattice variables, assuming that the electronic subsystem rapidly relaxes to the quasi-equilibrium state associated with the instantaneous configuration $\{Q_i\}$. The lattice coordinates then obey the Langevin equation
\begin{equation}
M\frac{d^2Q_i}{dt^2}
=
F_i
-\gamma\frac{dQ_i}{dt}
+\eta_i(t),
\label{eq:langevin}
\end{equation}
where $\gamma$ is the damping coefficient and $\eta_i(t)$ is a Gaussian thermal noise satisfying
\begin{equation}
\langle\eta_i(t)\eta_j(t')\rangle
=
2\gamma k_{\rm B}T\,\delta_{ij}\delta(t-t').
\label{eq:fdt}
\end{equation}
The deterministic force contains a classical elastic contribution and an electronic contribution. For the full semiclassical model, the latter is obtained from the instantaneous electronic state through the Hellmann--Feynman theorem, giving
\begin{equation}
F_i
=
-KQ_i
-\kappa\sum_{j\in{\cal N}(i)}Q_j
+
g\left(\langle n_i\rangle-\frac{1}{2}\right),
\label{eq:full_force}
\end{equation}
where ${\cal N}(i)$ denotes the four nearest neighbors of site $i$.

The difficulty lies in evaluating $\langle n_i\rangle$. For every instantaneous lattice configuration, one must solve the itinerant single-particle Hamiltonian, determine the Fermi occupations, and adjust the chemical potential to maintain half filling. Repeating this electronic calculation at every Langevin time step becomes prohibitively expensive for the large two-dimensional lattices and long simulation times needed to study coarsening. This computational bottleneck motivated the strong-coupling reduction introduced in our previous work~\cite{Jang2026}.

We therefore consider the quasiclassical limit of the semiclassical Holstein model. The relevant hierarchy of scales is
\begin{equation}
\hbar\Omega_0 \ll t_{\rm nn}\ll gQ_0 ,
\label{eq:qcl_hierarchy}
\end{equation}
where $\Omega_0=\sqrt{K/M}$ is the characteristic lattice frequency and $Q_0\sim g/K$ is the characteristic distortion amplitude. The first inequality expresses the adiabatic separation between the slow lattice and fast electronic dynamics, while the second places the electronic problem in the strong-coupling regime. Electron hopping may then be treated perturbatively in the small parameter $t_{\rm nn}/(gQ_0)$~\cite{Jang2026}.

It is important that this approximation not be interpreted as simply defining a model with $t_{\rm nn}=0$. Rather, the quasiclassical limit corresponds to the leading term of a strong-coupling expansion of the electronic contribution to the lattice force. To this order, the electronic Hamiltonian becomes local,
\begin{equation}
\mathcal{H}_{\rm e}^{(0)}
=
-g\sum_i Q_i n_i ,
\label{eq:qcl_electronic}
\end{equation}
while hopping processes enter only through higher-order corrections.

For a fixed lattice configuration, the electronic eigenstates of Eq.~\eqref{eq:qcl_electronic} are simply local occupation-number states. The single-particle energy of site $i$ is $\epsilon_i=-gQ_i$, and the quasi-equilibrium occupation is therefore
\begin{equation}
\langle n_i\rangle
=
\frac{1}
{\exp[-(gQ_i+\mu)/T]+1},
\label{eq:qcl_occupation}
\end{equation}
where the chemical potential $\mu$ is determined at every time step from the global half-filling constraint
\begin{equation}
\sum_i \langle n_i\rangle=\frac{N}{2}.
\label{eq:half_filling}
\end{equation}
Although the electronic Hamiltonian has become local, the fermionic statistics and global particle-number constraint are retained. This mixture of classical lattice coordinates and Fermi--Dirac electronic occupations is the reason for referring to the approximation as quasiclassical~\cite{Jang2026}.

At zero temperature, Eq.~\eqref{eq:qcl_occupation} approaches a binary occupation pattern: the $N/2$ sites with the lowest instantaneous electronic energies are occupied and the remainder are empty. At finite temperature, the Fermi edge is broadened and sites whose local energies lie near the chemical potential acquire partial occupations. This distinction will become important below, since the degree to which the occupations are binary directly controls the charge constraint on local domain-wall motion.

The purely local electronic problem by itself does not retain the intersite coupling responsible for selecting the checkerboard pattern. We therefore keep the nearest-neighbor elastic interaction in Eq.~\eqref{eq:holstein}, which may be viewed either as a direct cooperative lattice interaction or as an effective coupling representing higher-order virtual hopping processes. The quasiclassical Hamiltonian used in the simulations is consequently
\begin{equation}
\mathcal{H}_{\rm qcl}
=
\sum_i\left(\frac{P_i^2}{2M}+\frac{K}{2}Q_i^2\right)
+\kappa\sum_{\langle ij\rangle}Q_iQ_j
-g\sum_i\left(n_i-\frac{1}{2}\right)Q_i .
\label{eq:qcl_hamiltonian}
\end{equation}
For $\kappa>0$, the intersite term favors staggered distortions and stabilizes the $(\pi,\pi)$ CDW state. In the simulations below, $\kappa$ is treated as an effective model parameter rather than being fixed microscopically by $t_{\rm nn}$.

A particularly useful feature of the quasiclassical limit is that the electronic force can now be written explicitly. Using Eq.~\eqref{eq:qcl_occupation},
\begin{equation}
F_i^{\rm qcl}
=
-KQ_i
-\kappa\sum_{j\in{\cal N}(i)}Q_j
+
\frac{g}{2}
\tanh\left(\frac{gQ_i+\mu}{2T}\right).
\label{eq:qcl_force}
\end{equation}
Equations~\eqref{eq:langevin}, \eqref{eq:half_filling}, and \eqref{eq:qcl_force} define the dynamical problem studied in this work. At every time step, the chemical potential is adjusted to satisfy the total particle-number constraint, after which all local electronic forces follow directly from Eq.~\eqref{eq:qcl_force}. No diagonalization of an itinerant electronic Hamiltonian is required.

This reduction is crucial for the present two-dimensional study. The force evaluation is essentially local apart from determining the single global chemical potential, allowing lattices and evolution times far beyond those accessible with repeated exact diagonalization. At the same time, the approximation retains the two ingredients central to the physics discussed below: Fermi--Dirac occupations and conservation of the total electronic charge. The resulting dynamics therefore provides a simple setting in which the consequences of microscopic charge conservation for the coarsening of an otherwise nonconserved CDW order parameter can be isolated.

\section{Thermal-quench simulations and domain growth}
\label{sec:coarsening}

Having established the quasiclassical limit, we now use it to study the phase-ordering dynamics of the two-dimensional CDW following a thermal quench. The closed-form electronic force makes it possible to perform Langevin simulations on lattices sufficiently large to follow the development and growth of extended CDW domains over a broad range of times.

\subsection{Simulation protocol}

The equations of motion are integrated numerically using a scheme in which the damping and stochastic terms are treated exactly over each time step through the Ornstein--Uhlenbeck propagator, followed by a trapezoidal update of the lattice displacement. At every time step, the local electronic occupations are recomputed from the instantaneous lattice configuration using Eq.~\eqref{eq:qcl_occupation}, while the chemical potential is determined iteratively so as to enforce the half-filling condition in Eq.~\eqref{eq:half_filling}. Periodic boundary conditions are imposed in both spatial directions.

The simulations follow a thermal-quench protocol. We first prepare the system in a disordered state by equilibrating at a high temperature $T_{\rm high}=5$, well above the CDW ordering regime. The bath temperature is then suddenly lowered to a final temperature $T$ in the ordered phase, after which the system is evolved for a time $t_{\rm max}$. By varying the final temperature, we examine how the coarsening dynamics changes across the ordered phase.

Unless otherwise stated, simulations are performed on square lattices with $L_x=L_y=200$, corresponding to $N=4\times 10^4$ sites. The dimensionless model parameters are $\lambda=0.2$ and $\bar{\gamma}=0.2$, and the equations of motion are integrated with a time step $\Delta t=0.005$. The post-quench dynamics is followed up to $t_{\rm max}=1500$, with configurations recorded at intervals of $\Delta t_{\rm rec}=250$. For each final temperature, observables are averaged over 60 independent realizations.

\subsection{CDW domains and correlation functions}

\begin{figure}[t]
\centering
\safeincludegraphics[width=\columnwidth]{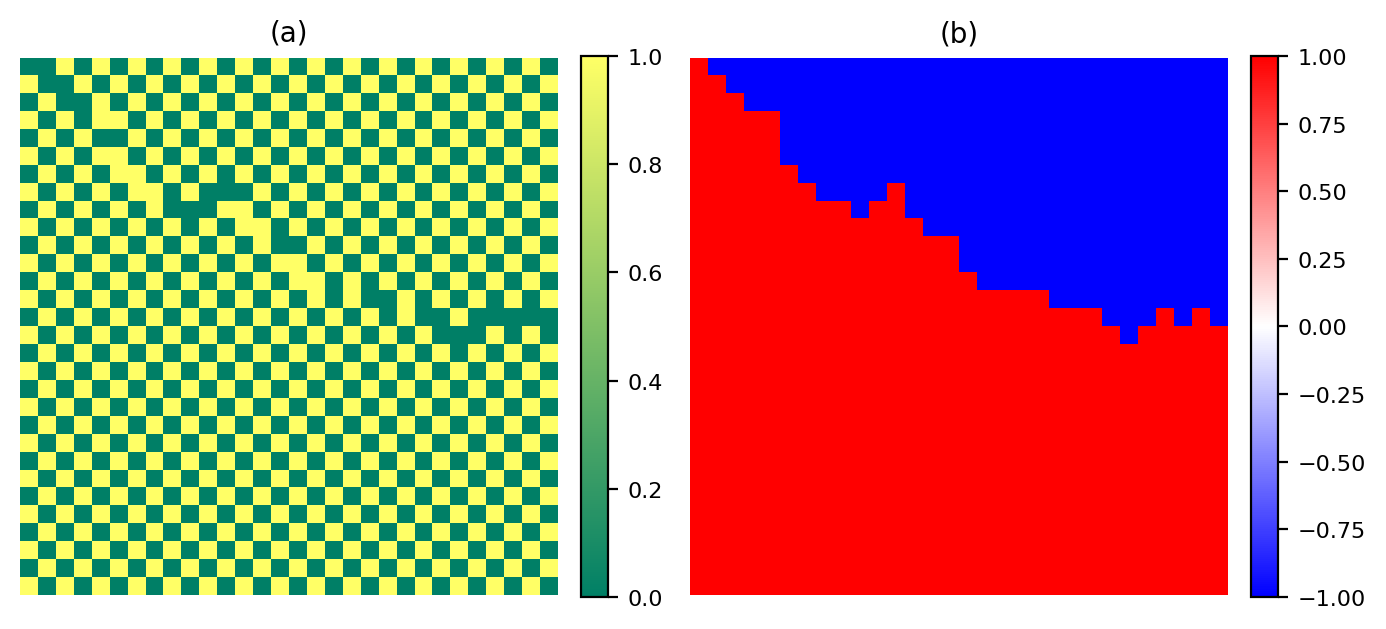}
\caption{
Relation between the microscopic charge pattern and the local CDW order parameter. (a) Representative checkerboard charge configuration $n_i$. (b) Corresponding staggered field $\phi_i$, which removes the microscopic checkerboard modulation and maps the two symmetry-related CDW states onto positive and negative domains.
}
\label{fig:charge_cdw}
\end{figure}

Figure~\ref{fig:charge_cdw} illustrates the relation between the microscopic charge configuration and the local CDW order parameter used throughout the analysis. Panel (a) shows a representative checkerboard charge configuration. While the alternating pattern of charge-rich and charge-poor sites directly reflects the microscopic CDW, it obscures the larger-scale domain structure. We therefore define a local staggered order parameter
\begin{equation}
\phi_i
=
(-1)^{x_i+y_i}
\left(
n_i-\frac{1}{2}
\right),
\label{eq:local_cdw}
\end{equation}
where $(x_i,y_i)$ are the integer coordinates of site $i$. As shown in Fig.~\ref{fig:charge_cdw}(b), this transformation removes the microscopic checkerboard modulation and maps the two symmetry-related CDW states onto domains with opposite signs of $\phi_i$.

\begin{figure*}[t]
\centering
\safeincludegraphics[width=0.95\textwidth]{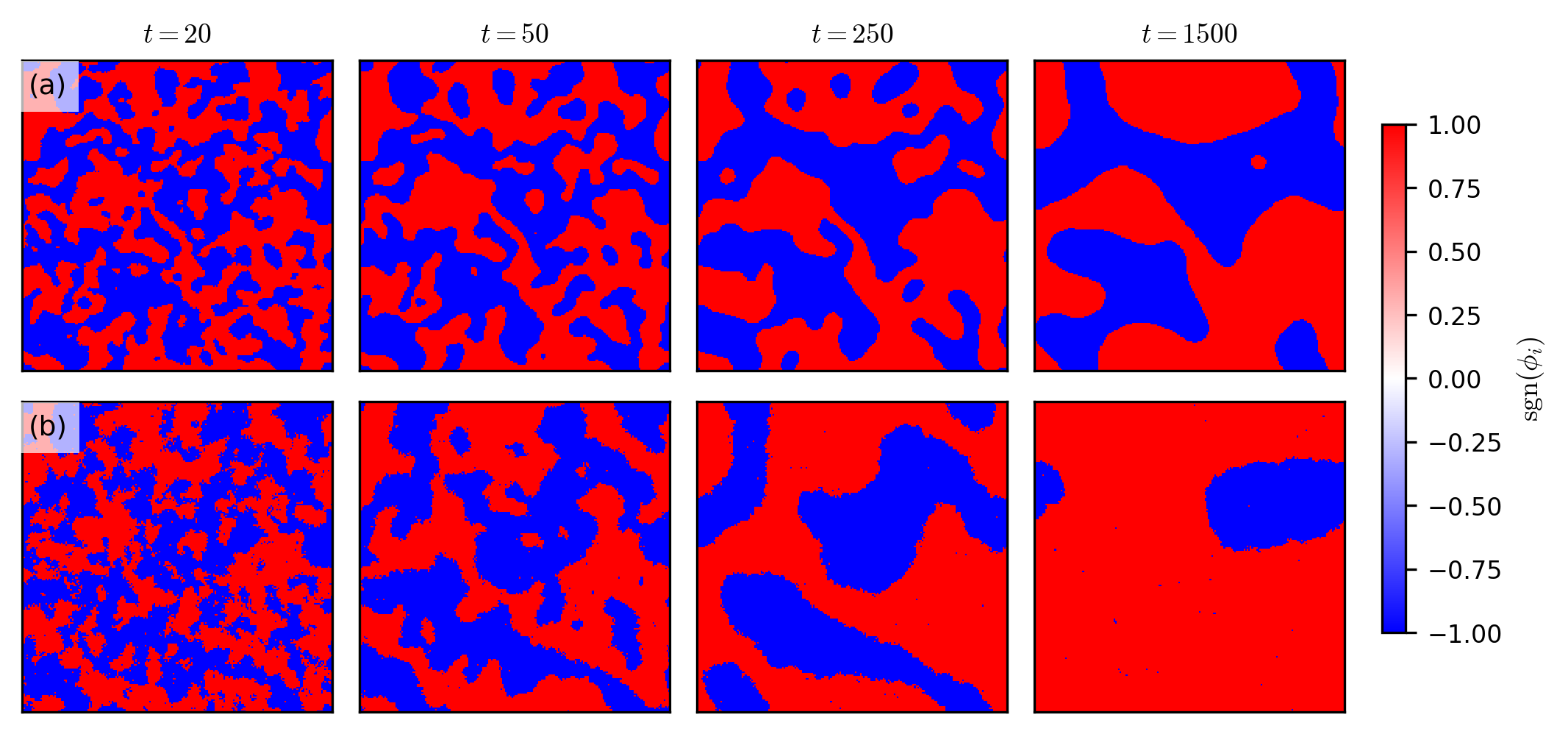}
\caption{
Evolution of CDW domains following thermal quenches on a $200\times200$ lattice. Colors indicate the sign of the local staggered CDW order parameter, $\mathrm{sgn}(\phi_i)$. (a) Quench to $T=0.05$. (b) Quench to $T=0.40$. In both cases, initially disordered domain patterns coarsen through the elimination of small domains and motion of the domain walls. The higher-temperature quench exhibits noticeably faster domain growth over the time window shown.
}
\label{fig:snapshots}
\end{figure*}

Representative real-space configurations during the post-quench evolution are shown in Fig.~\ref{fig:snapshots}. The colors denote the sign of $\phi_i$ and therefore directly visualize the evolving CDW domains. Shortly after the quench, the system consists of a dense network of small domains separated by irregular interfaces. As time increases, short-wavelength structures disappear, neighboring regions merge, and the characteristic domain size grows. Although the overall morphology is qualitatively similar at different final temperatures, the rate of coarsening changes substantially. For the low-temperature quench in Fig.~\ref{fig:snapshots}(a), a relatively dense domain-wall network persists over long times, whereas the higher-temperature quench in Fig.~\ref{fig:snapshots}(b) exhibits more rapid elimination of small domains and reaches substantially larger characteristic length scales at comparable times. This pronounced temperature dependence provides the first indication that the CDW coarsening kinetics changes significantly across the ordered phase and will be quantified below.

To characterize this growth quantitatively, we compute equal-time spatial correlation functions for both the microscopic lattice distortion $Q_i$ and the staggered CDW field $\phi_i$,
\begin{align}
C_Q(r,t)
&=
\frac{
\left\langle Q_i(t)Q_j(t)\right\rangle_{|\mathbf{r}_i-\mathbf{r}_j|=r}
}{
\left\langle Q_i^2(t)\right\rangle
},
\nonumber\\
C_\phi(r,t)
&=
\frac{
\left\langle \phi_i(t)\phi_j(t)\right\rangle_{|\mathbf{r}_i-\mathbf{r}_j|=r}
}{
\left\langle \phi_i^2(t)\right\rangle
}.
\label{eq:correlation_functions}
\end{align}
The averages are taken over lattice positions, angular directions of the separation vector, and independent realizations. The normalization is chosen such that $C_Q(0,t)=C_\phi(0,t)=1$. Considering both quantities is useful because they emphasize different spatial structures: $C_Q$ retains the microscopic checkerboard modulation of the underlying lattice distortion, whereas $C_\phi$ removes this rapid staggered variation and isolates the slowly varying CDW domain pattern.

\begin{figure}[t]
\centering
\safeincludegraphics[width=\columnwidth]{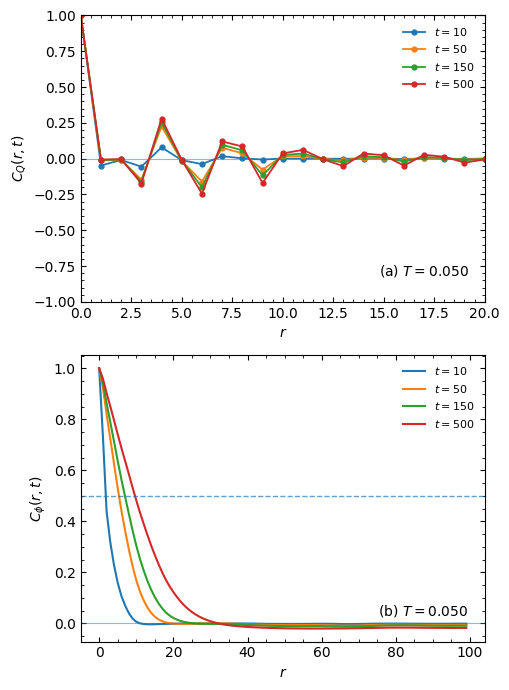}
\caption{
Equal-time correlation functions following a thermal quench to $T=0.050$. (a) Correlation function $C_Q(r,t)$ of the microscopic lattice distortion field. The alternating short-distance structure reflects the underlying checkerboard CDW modulation, while the spatial extent of its envelope grows with time. (b) Correlation function $C_\phi(r,t)$ of the staggered CDW order parameter. The dashed horizontal line marks the half-height criterion $C_\phi(r,t)=1/2$ used to define the characteristic domain length $L(t)$.
}
\label{fig:correlation_functions}
\end{figure}

This distinction is illustrated in Fig.~\ref{fig:correlation_functions} for a quench to $T=0.050$. The microscopic correlation function $C_Q(r,t)$ in panel (a) exhibits alternating signs at short distances as a direct consequence of the underlying $(\pi,\pi)$ checkerboard order. The oscillation remains locked to the lattice scale as the system evolves, while its envelope extends progressively to larger distances as the ordered domains grow. By contrast, $C_\phi(r,t)$ in panel (b) is smooth on the lattice scale and directly resolves the correlations between the two symmetry-related CDW domains. Its systematic broadening with time therefore provides a natural measure of the coarsening length scale. We define the characteristic domain size $L(t)$ through the half-height condition $C_\phi(L(t),t)=1/2$, using linear interpolation between neighboring radial bins. This construction removes the microscopic checkerboard structure and yields a convenient measure of the typical domain size for the growth analysis below.

\subsection{Domain growth and apparent power-law behavior}

For a scalar nonconserved order parameter, the standard phenomenological description of phase-ordering dynamics is the time-dependent Ginzburg--Landau equation, corresponding to Model A in the Hohenberg--Halperin classification~\cite{Hohenberg1977,Bray1994,Onuki2002,SanjayPuri2009}. The order parameter relaxes locally toward lower free energy without an associated conservation law,
\begin{equation}
\frac{\partial \phi(\mathbf{r},t)}{\partial t}
=
-\Gamma
\frac{\delta \mathcal{F}}{\delta \phi(\mathbf{r},t)},
\label{eq:model_A}
\end{equation}
where $\Gamma$ is a kinetic coefficient.
For an Ising-like scalar field, the coarse-grained free energy may be taken in the standard Ginzburg--Landau form
\begin{equation}
\mathcal{F}[\phi]
=
\int d^2r\,
\left[
\frac{\kappa_\phi}{2}|\nabla\phi|^2
+
\frac{a}{2}\phi^2
+
\frac{b}{4}\phi^4
\right],
\label{eq:GL_free_energy}
\end{equation}
with $a<0$ and $b>0$ in the ordered phase. The two minima of the local potential correspond to the two symmetry-related Ising states, while the gradient term assigns an energetic cost to domain walls separating them.

The resulting dynamics is purely relaxational: interfaces evolve so as to reduce the coarse-grained free energy, and at late times their motion is governed by local curvature. For a domain of characteristic size $L$, the typical curvature scales as $1/L$, so that the characteristic growth rate decreases as domains become larger. This leads to the Allen--Cahn growth law $L(t)\sim t^{1/2}$ for a nonconserved scalar order parameter in two or more dimensions~\cite{Allen1972,Bray1994}. In the conventional scaling description introduced above, this corresponds to the dynamical exponent $z=2$, and therefore provides the natural reference behavior for the Ising-like CDW order studied here.

\begin{figure}[t]
\centering
\safeincludegraphics[width=0.95\columnwidth]{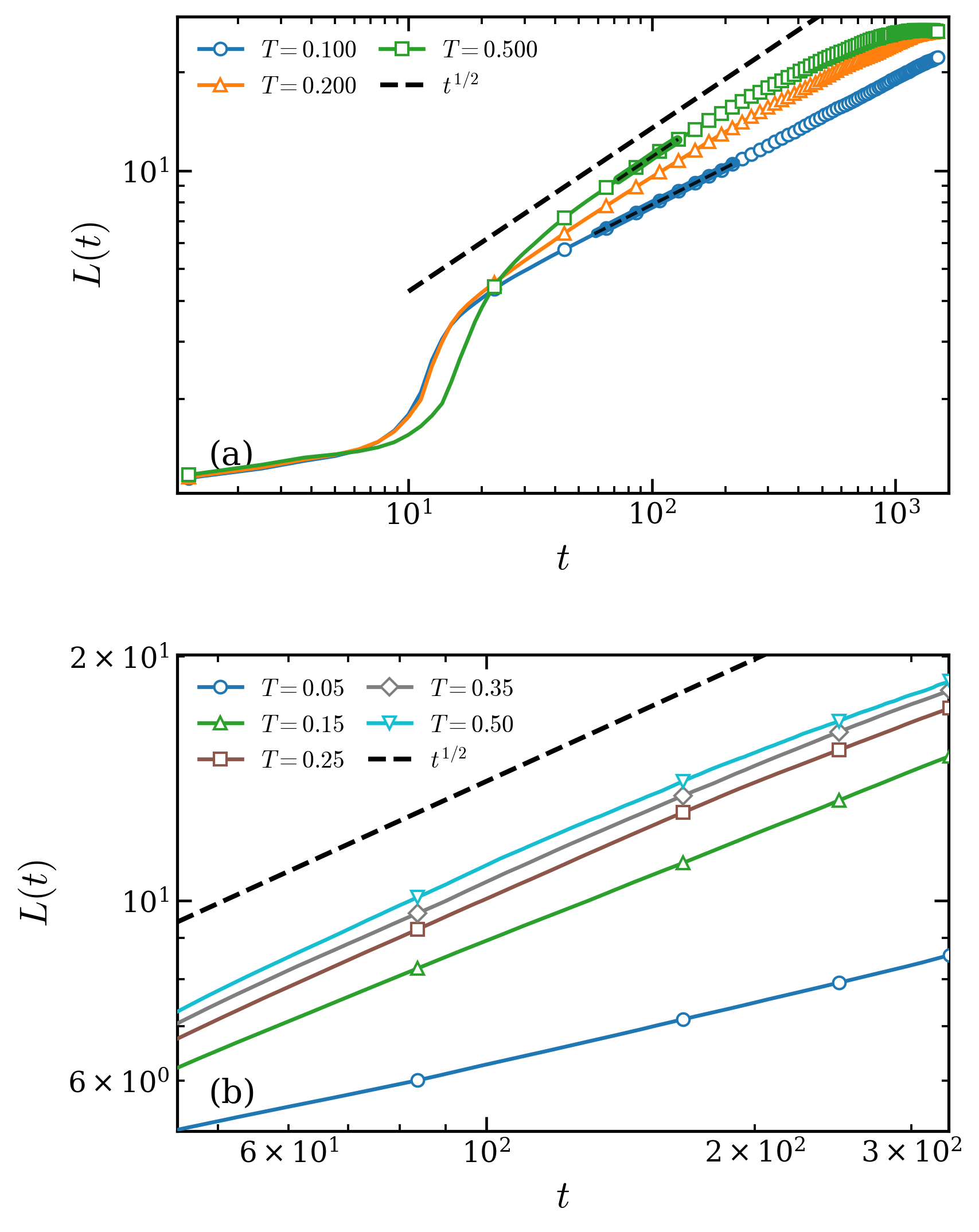}
\caption{
Characteristic domain length following thermal quenches.
(a) Log--log plot of $L(t)$ for representative final temperatures.
The dashed line indicates the Allen--Cahn reference behavior,
$L(t)\sim t^{1/2}$.
(b) Enlarged view of the same regime for a broader set of temperatures,
showing the systematic increase of the apparent growth rate with temperature.
For the representative curves shown in panel (b), the apparent growth
exponents are extracted over the time intervals
$t\in[80,150]$ for $T=0.05$,
$[60,180]$ for $T=0.15$,
$[110,200]$ for $T=0.25$,
$[80,180]$ for $T=0.35$, and
$[65,135]$ for $T=0.50$.
The precise fitting window varies modestly with temperature and is chosen
within the intermediate-time coarsening regime where $L(t)$ exhibits
approximately power-law behavior.
}
\label{fig:length_growth}
\end{figure}

Motivated by this expectation, we examine the time dependence of the characteristic length extracted from $C_\phi(r,t)$. Figure~\ref{fig:length_growth}(a) shows $L(t)$ on logarithmic axes for several representative quench temperatures. Following an initial transient associated with the formation of well-developed local CDW order, the growth curves enter extended time intervals that appear approximately algebraic. The dashed line indicates the Allen--Cahn form $L(t)\sim t^{1/2}$ for comparison. The numerical growth is substantially slower at low temperature, while increasing the final temperature systematically steepens the curves and brings them progressively closer to the square-root law. Figure~\ref{fig:length_growth}(b) enlarges the late-time portion of the simulations and makes this temperature dependence more apparent.

Because the simulations probe a finite time window, we characterize the observed growth empirically by fitting the late-time data to
%\begin{equation}
$L(t)\sim t^{\alpha},$
%\label{eq:apparent_growth}
%\end{equation}
where $\alpha$ should be understood as an apparent growth exponent extracted over a restricted time interval, rather than identified a priori with the asymptotic exponent $1/z$. The dashed guide lines on top of L(t) in Fig.~\ref{fig:length_growth} are drawn over representative intermediate-time ranges that are comparable to the windows used to extract the apparent exponents for the different temperatures. The precise fitting windows vary slightly with temperature, but all lie within this intermediate coarsening regime. This distinction is important here because the numerical curves need not have reached a true asymptotic scaling regime over the accessible simulation times.

\begin{figure}[t]
\centering
\safeincludegraphics[width=0.95\columnwidth]{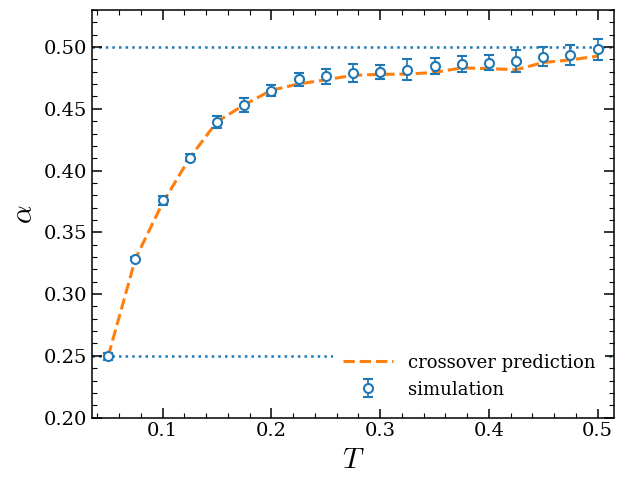}
\caption{
Apparent coarsening exponent $\alpha(T)$ obtained from power-law fits to $L(t)$ over the intermediate-time growth regime. The dotted horizontal lines indicate $\alpha=1/4$ and the Allen--Cahn value $\alpha=1/2$. The orange dashed curve shows the prediction obtained from the direct-rate analysis developed in the following section. Error bars indicate the sensitivity of the fitted exponent to the fitting window.
}
\label{fig:alpha_vs_T}
\end{figure}

The resulting temperature dependence of the fitted exponent is summarized in Fig.~\ref{fig:alpha_vs_T}. At the lowest temperatures, the measured $\alpha$ is close to $1/4$. It then increases continuously with temperature and approaches the Allen--Cahn value $1/2$ at the upper end of the range studied. We emphasize that the quantity plotted in Fig.~\ref{fig:alpha_vs_T} is therefore a finite-time, apparent exponent determined from the late-time fitting windows of Fig.~\ref{fig:length_growth}, rather than a directly measured asymptotic dynamical exponent. The origin of this continuous evolution, and the reason the two limiting values $1/4$ and $1/2$ emerge naturally, will be developed in the following section.

\section{Cooperative domain-wall motion and coarsening crossover}
\label{sec:coarsening_mechanism}

\begin{figure*}[t]
    \centering
    \safeincludegraphics[width=\textwidth]{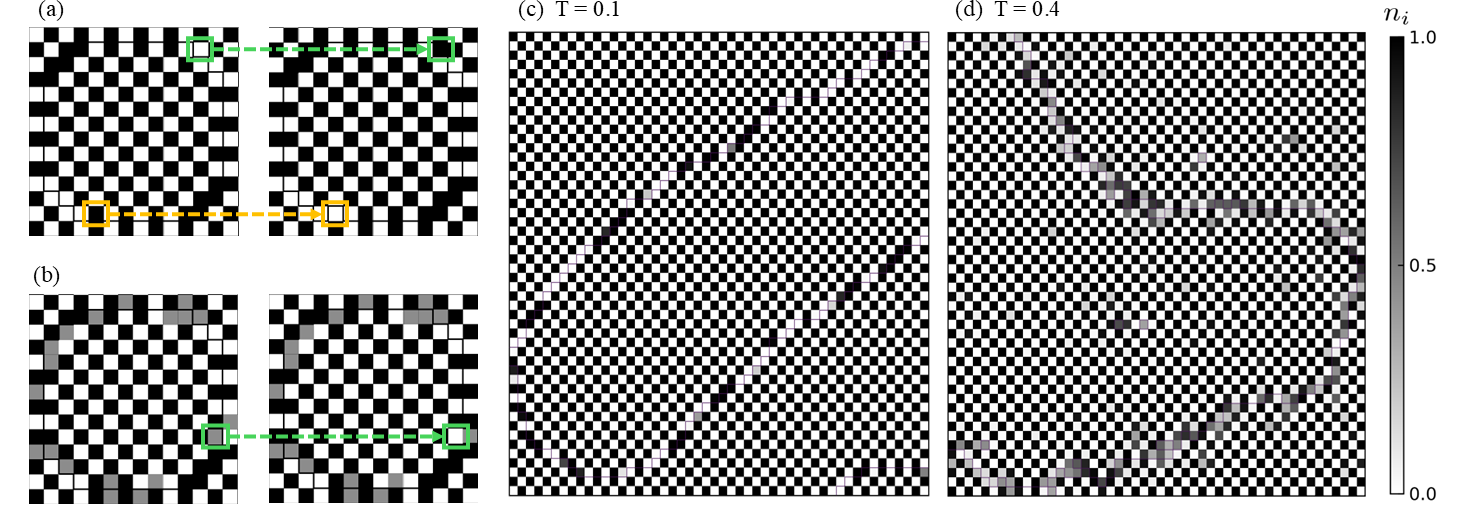}
    \caption{Schematic illustration of the two mechanisms of domain-wall motion and their temperature dependence, based on snapshots from the simulations. Black and white squares denote occupied $(n_i=1)$ and empty $(n_i=0)$ sites, respectively, while gray squares indicate intermediate occupation. (a) Cooperative wall motion: two spatially separated wall segments rearrange simultaneously so that the total electron number is conserved. The colored boxes identify the participating segments, and the dotted arrows indicate their displacements. (b) Fuzzy-wall motion: a partially occupied boundary site permits a local rearrangement without requiring a simultaneous displacement of a spatially separated wall segment. (c,d) Local occupation profiles $n_i$ at $t=500$ for $T=0.1$ and $0.4$, respectively. The thin purple curves mark the boundaries between the two checkerboard CDW domains. Intermediate occupations are relatively scarce at low temperature but become prevalent along the domain walls at higher temperature, opening additional channels for local wall motion.}
    \label{fig:schematic}
\end{figure*}

The apparent growth exponents in Fig.~\ref{fig:alpha_vs_T} lie approximately between $\alpha=1/4$ and $1/2$. These two limiting values suggest distinct kinetic regimes of CDW domain-wall motion. The upper limit corresponds to conventional Allen--Cahn coarsening, whereas the lower limit can be understood through a cooperative mechanism constrained by conservation of the electronic charge. We first discuss the physical picture and scaling arguments underlying these two limits.

For a nonconserved scalar order parameter, the Allen--Cahn description states that the local normal speed of an interface is proportional to its curvature, $v\propto\kappa_{\mathrm{dw}}$, with motion directed toward reducing the interfacial energy. In a coarsening state characterized by a single length scale $L(t)$, the typical curvature scales as $1/L$. For a size-independent interface mobility, the characteristic growth rate and the resulting growth law are therefore
\begin{equation}
\frac{dL}{dt}\sim\frac{D(T)}{L},
\qquad
L(t)\sim[D(T)t]^{1/2},
\label{eq:allen_cahn_growth}
\end{equation}
where $D(T)$ is an effective kinetic coefficient that incorporates the interface mobility, interfacial tension, and geometric factors. This gives the conventional Allen--Cahn growth exponent $\alpha=1/2$, corresponding to the dynamical exponent $z=2$.

The slower limiting behavior is motivated by the cooperative mechanism identified in our previous study of the one-dimensional Holstein model~\cite{Jang2026}. In one dimension, CDW domains are separated by localized kinks. Displacing an individual kink changes the electronic occupation near the defect and therefore requires a compensating rearrangement elsewhere to conserve the total electron number. The motion of a kink consequently depends on the availability of a compatible partner. In two dimensions, the defects are extended domain walls rather than isolated kinks, but the same charge-compensation requirement can constrain local interface motion: a rearrangement that increases the occupation near one wall segment must be accompanied by another that decreases it elsewhere.

This constraint is particularly important at low temperature, where the electronic occupations are nearly binary. Two compatible wall segments can then rearrange cooperatively, allowing the interface to move while conserving the total charge, as illustrated in Fig.~\ref{fig:schematic}(a). To describe the availability of such rearrangements, we introduce the domain-wall density $\rho_{\mathrm{dw}}$, defined as the total interface length per unit area. A simple factorized estimate assigns a statistical weight proportional to $\rho_{\mathrm{dw}}^2$ to pairs of interfacial regions, with their local compatibility incorporated into a temperature-dependent prefactor. Motivated by this pair requirement, we consider a scaling description in which the effective wall mobility is suppressed by the same density factor. The resulting growth equation is
\begin{equation}
\frac{dL}{dt}\sim\frac{D_{\rm pair}(T)\rho_{\mathrm{dw}}^2}{L},
\label{eq:cooperative_growth_density}
\end{equation}
where $D_{\rm pair}(T)$ includes the microscopic rate and length factors associated with cooperative rearrangements. The density dependence in this equation is a kinetic assumption that will be tested against the measured growth rates below.

In the dynamical scaling regime, the number of domains per unit area scales as $L^{-2}$, while the typical domain perimeter scales as $L$. Their product therefore gives $\rho_{\mathrm{dw}}\sim L^{-1}$. Substituting this relation into Eq.~\eqref{eq:cooperative_growth_density}, with geometric factors absorbed into $D_{\rm pair}(T)$, yields
\begin{equation}
\frac{dL}{dt}\sim\frac{D_{\rm pair}(T)}{L^3},
\qquad
L(t)\sim[D_{\rm pair}(T)t]^{1/4}.
\label{eq:cooperative_growth}
\end{equation}
Within this description, curvature continues to drive domain growth, but the decreasing availability of cooperative rearrangements reduces the effective wall mobility as the interfaces become dilute. This additional kinetic suppression produces the limiting growth exponent $\alpha=1/4$, corresponding to the dynamical exponent $z=4$.

The remaining question is how conventional Allen--Cahn growth is recovered as the temperature increases. Deep in the CDW phase, the electronic occupations are close to either zero or one, and a local displacement of a domain wall involves an almost discrete change in the occupation pattern. Conservation of the total electron number then strongly constrains individual wall rearrangements, favoring the cooperative mechanism illustrated in Fig.~\ref{fig:schematic}(a). At finite temperature, however, thermal broadening allows intermediate electronic occupations, particularly near domain walls where the local CDW order is weakened. These partially occupied sites provide additional possibilities for redistributing charge during interface motion. The cooperative constraint therefore becomes less restrictive as the temperature rises, allowing local wall motion that does not require a second, discrete compensating wall displacement.

Figure~\ref{fig:schematic}(b) illustrates this alternative process through the motion of ``fuzzy'' boundary sites with occupations near $n_i=1/2$. Unlike the nearly binary sites in the low-temperature configuration, these sites can change their occupations continuously as the local lattice distortions evolve. A neighboring site can become partially occupied while another develops a more nearly integer occupation, allowing the diffuse boundary region to propagate through local updates. Such motion remains consistent with global charge conservation through continuous charge redistribution, but it does not require the simultaneous rearrangement of two distinct wall segments characteristic of the cooperative process. Consequently, this local channel can sustain a wall mobility without the additional pair-availability factor invoked in Eq.~\eqref{eq:cooperative_growth_density}. Together with the curvature driving force, it provides a route to the Allen--Cahn growth law.

The occupation snapshots in Fig.~\ref{fig:schematic}(c) and (d) provide a microscopic illustration of this temperature dependence. At $T=0.1$, the configuration consists predominantly of nearly occupied and empty sites, shown in black and white, respectively. The checkerboard occupation pattern remains sharply defined up to the vicinity of the domain boundaries, and relatively few sites exhibit intermediate occupations. At $T=0.4$, intermediate gray values are substantially more prominent along the boundaries marked by the purple curves, indicating an increased population of partially occupied interfacial sites. These configurations support the physical picture that thermal broadening opens additional local channels for domain-wall motion. The relative importance of cooperative rearrangements and local motion through partially occupied sites can therefore change with temperature, motivating a crossover description that includes both contributions to the growth rate.

\begin{figure}[t]
    \centering
    \safeincludegraphics[width=\linewidth]{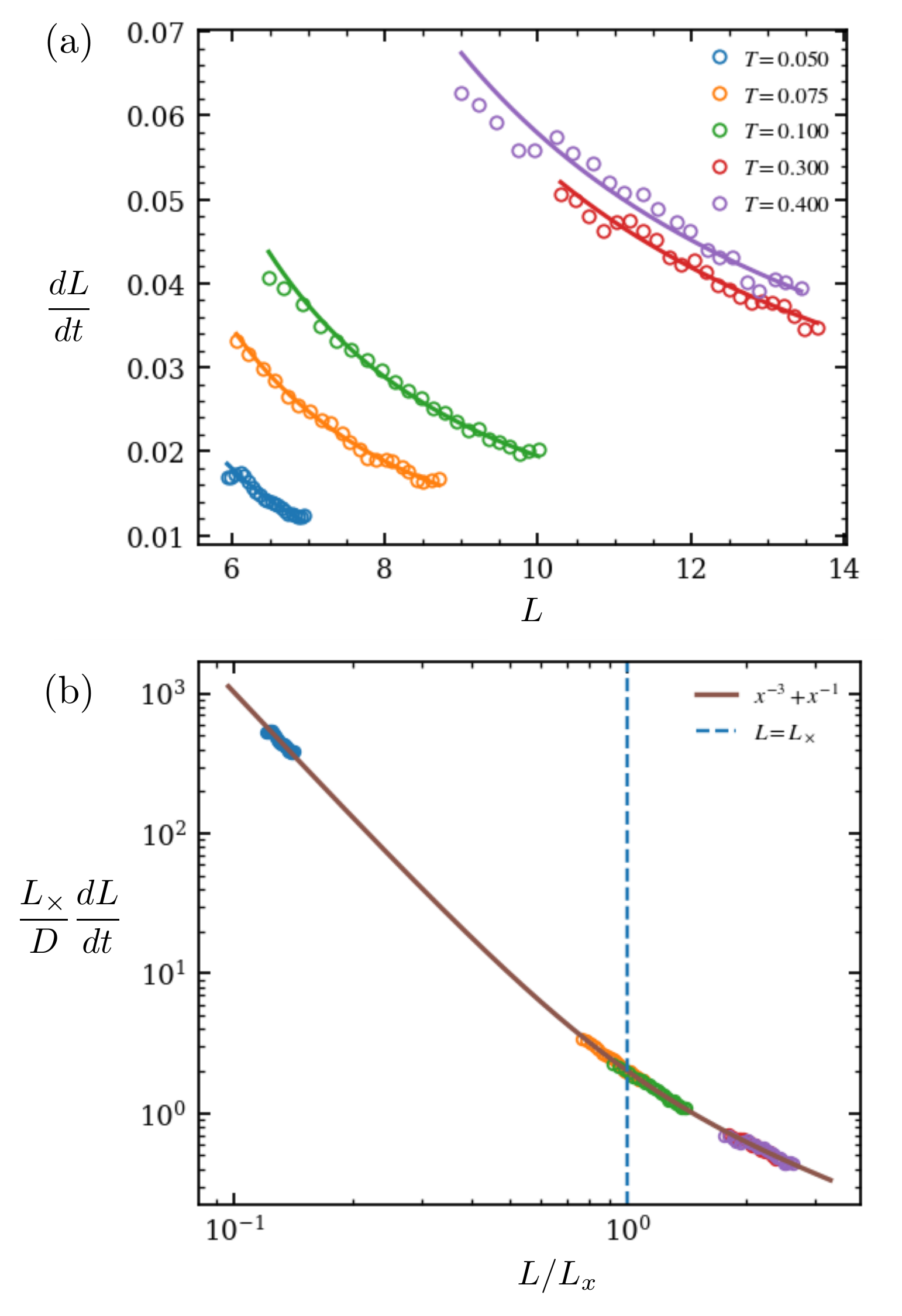}
    \caption{Direct test of the coarsening crossover model and the resulting scaling collapse. (a) Measured growth rate $\dot{L}$ as a function of the instantaneous domain size $L$ for selected temperatures. Open circles show the simulation results, while solid curves show $\dot{L}=D(T)/L+D_{\rm pair}(T)/L^3$ using the extracted coefficients $D(T)$ and $D_{\rm pair}(T)$. (b) The same data rescaled using the crossover length $L_{\times}=\sqrt{D_{\rm pair}/D}$. The solid curve is the dimensionless prediction $(L_{\times}/D)\dot{L}=x^{-1}+x^{-3}$, where $x=L/L_{\times}$. Colors denote the same temperatures as in panel (a), and the dashed line at $x=1$ marks equal contributions from cooperative and local domain-wall motion.}
    \label{fig:Lt-analysis}
\end{figure}

At finite temperature, both cooperative rearrangements and local motion through partially occupied sites can contribute to the evolution of the domain walls. A minimal description treats them as parallel kinetic channels, with respective contributions $D(T)/L$ and $D_{\rm pair}(T)/L^3$ to the growth rate. Defining the crossover length $L_\times(T)\equiv[D_{\rm pair}(T)/D(T)]^{1/2}$, their sum can be written compactly as
\begin{equation}
\frac{dL}{dt}=\frac{D(T)}{L}\left[1+\frac{L_\times^2(T)}{L^2}\right].
\label{eq:crossover_growth}
\end{equation}
The first contribution describes conventional curvature-driven motion enabled by local charge redistribution through partially occupied sites. For the cooperative contribution, the curvature driving force again supplies a factor $1/L$, while the probability of finding two compatible interfacial regions introduces the additional factor $\rho_{\mathrm{dw}}^2\sim 1/L^2$. The coefficients $D(T)$ and $D_{\rm pair}(T)$ incorporate the microscopic rates and temperature-dependent statistical weights of the two processes; they are kinetic coefficients rather than normalized probabilities. By construction, the two contributions are equal at $L=L_\times$.

At fixed temperature, Eq.~\eqref{eq:crossover_growth} can be integrated exactly. Suppressing the explicit temperature dependence of $D$ and $L_\times$, and writing $L_0=L(t_0)$, one obtains the implicit solution
\begin{equation}
2D(t-t_0)=L^2-L_0^2-L_\times^2
\ln\left(\frac{L^2+L_\times^2}{L_0^2+L_\times^2}\right).
\label{eq:crossover_solution}
\end{equation}
For $L\ll L_\times$, expansion of the logarithm recovers $L(t)\sim[D_{\rm pair}(T)t]^{1/4}$, whereas for $L\gg L_\times$, the logarithmic correction becomes subleading to the quadratic term and the Allen--Cahn law $L(t)\sim[D(T)t]^{1/2}$ is recovered. Thus, provided $D(T)>0$, an extended cooperative regime ultimately crosses over to conventional Allen--Cahn growth. At low temperature, the scarcity of partially occupied sites strongly suppresses the local channel and pushes the crossover to larger length and time scales, potentially beyond the accessible simulation window.

The physical origin of this late-time crossover is the progressive dilution of domain walls. As $L$ grows, the wall density $\rho_{\mathrm{dw}}\sim L^{-1}$ decreases, reducing the statistical availability of compatible pairs of wall segments for cooperative motion. Local rearrangements through partially occupied sites require no separate compensating wall displacement and therefore avoid this additional suppression. Provided the local mobility remains nonzero, even a weak local channel eventually dominates, since the ratio of cooperative to local contributions decreases as $(L_\times/L)^2$. Consequently, for $L\gg L_\times$, the effective wall mobility becomes approximately independent of domain size, and curvature-driven motion recovers the Allen--Cahn law $L(t)\sim t^{1/2}$.

Between these two limits, the exact solution in Eq.~\eqref{eq:crossover_solution} is not a power law. An exponent obtained by fitting $L(t)$ over a restricted time interval therefore characterizes the portion of the crossover sampled by that interval rather than a distinct asymptotic growth law. The extracted crossover lengths are of order $L_\times\sim 10$ in lattice units for most of the temperatures studied but become substantially larger at the lowest temperature. Since the domain sizes reached in the simulations are comparable to these crossover scales, the measured curves naturally sample different stages of the crossover. The continuous temperature dependence of the apparent exponent $\alpha(T)$ in Fig.~\ref{fig:alpha_vs_T} can therefore be understood primarily as a crossover effect: the low-temperature curves remain closer to the cooperative regime, whereas the higher-temperature curves extend further into the Allen--Cahn regime.

To test this description quantitatively, Fig.~\ref{fig:Lt-analysis}(a) compares the measured growth rate $\dot{L}$ with Eq.~\eqref{eq:crossover_growth} as a function of the instantaneous domain size. For each temperature, a single pair of coefficients $D(T)$ and $D_{\rm pair}(T)$ describes both the overall magnitude and the size dependence of the growth rate over the displayed interval. The agreement extends from the slow growth at low temperature to the more rapid growth at higher temperature, supporting a common kinetic description of the different growth curves. Because this comparison tests $\dot{L}$ directly as a function of $L$, it does not require assigning a constant power-law exponent to the time-dependent domain size.

Figure~\ref{fig:Lt-analysis}(b) shows the same data after rescaling $L$ by $L_\times$ and $\dot{L}$ by the characteristic rate $D/L_\times$. With $x=L/L_\times$, Eq.~\eqref{eq:crossover_growth} predicts the temperature-independent master curve $x^{-1}+x^{-3}$. The rescaled data follow this common curve over the sampled range, with the lowest-temperature points lying well within the cooperative regime $x\ll 1$, the intermediate-temperature points near the crossover $x=1$, and the higher-temperature points in the regime where local Allen--Cahn motion becomes dominant. The dashed line at $x=1$ marks equal contributions from the two mechanisms. The collapse shows directly that the apparently different growth behaviors arise from sampling different portions of the same crossover and supports interpreting the temperature-dependent exponents as finite-window effective exponents.

The crossover model also quantitatively accounts for the apparent exponents summarized in Fig.~\ref{fig:alpha_vs_T}. For each temperature, we integrate Eq.~\eqref{eq:crossover_growth} using the coefficients $D(T)$ and $D_{\rm pair}(T)$ extracted from the direct-rate analysis, and then fit the resulting $L(t)$ over the same time window used for the simulation data. The apparent exponent obtained from this procedure is shown by the orange dashed curve. Its close agreement with the measured $\alpha(T)$ over the full temperature range is achieved without fitting the exponent itself, demonstrating that the continuous increase from values near $1/4$ toward $1/2$ follows quantitatively from the crossover between cooperative and local domain-wall motion.

\begin{figure}[t]
    \centering
    \safeincludegraphics[width=\columnwidth]{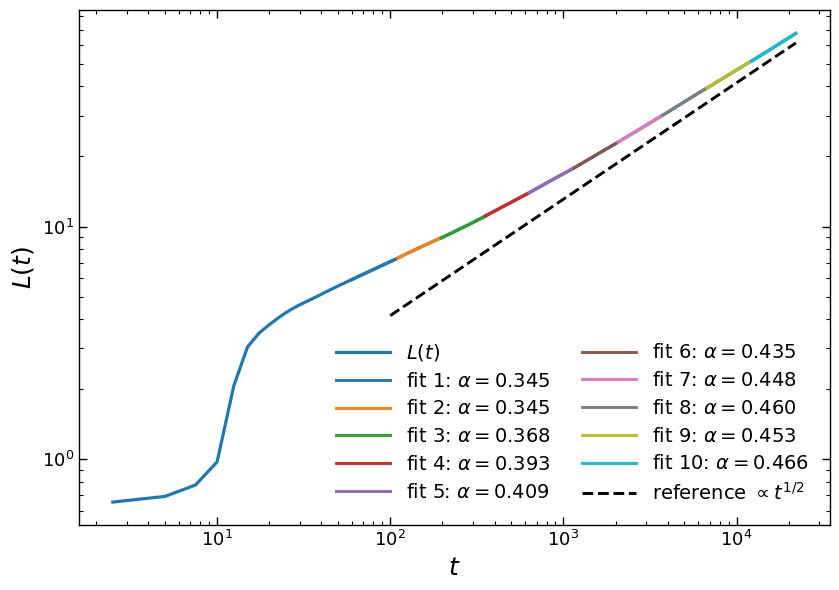}
\caption{Characteristic domain size $L(t)$ for a $500\times500$ system quenched to $T=0.075$. Colored segments show power-law fits over successive time windows, with the apparent exponents $\alpha$ listed in the legend. Their evolution from approximately $1/3$ toward $1/2$ reveals the coarsening crossover; the dashed line indicates the Allen--Cahn reference $L(t)\sim t^{1/2}$.}
\label{fig:Lt_crossover}
\end{figure}

Figure~\ref{fig:Lt_crossover} provides a direct time-domain demonstration of this crossover within a single coarsening trajectory. The large $500\times500$ system allows the growth at $T=0.075$ to be followed over a broad range of length and time scales while minimizing finite-size effects. Power-law fits over successive time windows show an overall increase of the apparent exponent from $\alpha\simeq 1/3$ at earlier times to values close to $1/2$ at late times. The modest fluctuations between neighboring windows reflect the uncertainty of local fits, but the systematic upward drift demonstrates that the trajectory is not governed by a single power law.

This evolution is precisely the behavior expected from Eq.~\eqref{eq:crossover_growth}. The ratio of the cooperative contribution to the local Allen--Cahn contribution is $L_\times^2/L^2$ and therefore decreases continuously as the domains grow. The earlier part of the trajectory remains influenced by cooperative pair motion and consequently exhibits an exponent below $1/2$, although the accessible window is already beyond the limiting $\alpha=1/4$ regime. At larger $L/L_\times$, local wall motion progressively dominates and the apparent exponent approaches the Allen--Cahn value. The figure thus illustrates explicitly how a sequence of seemingly well-defined effective exponents can emerge from different portions of a single smooth crossover.

\section{Conclusion and outlook}
\label{sec:conclusion}

We have studied CDW coarsening in the two-dimensional quasiclassical Holstein model, where the staggered CDW order parameter is nonconserved but its dynamics remains constrained by conservation of the microscopic electronic charge. Our results generalize the cooperative kink mechanism identified previously in one dimension~\cite{Jang2026} to extended domain walls. At low temperature, nearly binary electronic occupations require compatible wall segments to rearrange cooperatively, producing a scale-dependent mobility and the limiting growth law $L(t)\sim t^{1/4}$. Thermal broadening generates partially occupied interfacial sites that permit local wall motion and progressively restore Allen--Cahn growth, $L(t)\sim t^{1/2}$. The direct-rate analysis, scaling collapse, and long-time simulation on the $500\times500$ lattice consistently show that the measured temperature-dependent exponents describe different portions of this crossover rather than distinct asymptotic power laws.

The crossover description does not, however, exclude an additional temperature dependence of the cooperative dynamics itself. Partial electronic occupations could modify the density dependence of the cooperative mobility from $\rho_{\mathrm{dw}}^2$ to $\rho_{\mathrm{dw}}^{\zeta(T)}$, potentially contributing to a temperature-dependent correction to the apparent growth exponent in the regime $L\ll L_\times$. Within the same scaling argument, this would yield a cooperative growth exponent $\alpha_{\mathrm{coop}}(T)=1/[2+\zeta(T)]$, rather than a universal value of $1/4$. Provided $\zeta(T)>0$ and the local mobility remains nonzero, cooperative motion would still become subdominant at sufficiently large domain sizes, preserving the eventual crossover to Allen--Cahn growth. This possible influence of electronic populations on the cooperative exponent is not explored here and remains an open question.

The cooperative mechanism should persist beyond the quasiclassical limit when finite electron hopping is restored. Hopping provides additional channels for charge redistribution and can therefore modify the kinetic coefficients and crossover scale, but it does not remove the underlying requirement that domain-wall motion remain compatible with electron-number conservation. Indeed, anomalously slow CDW coarsening has already been observed in machine-learning simulations of the full two-dimensional semiclassical Holstein model with itinerant electrons~\cite{Cheng2023b}, suggesting that the behavior found here is not an artifact of the strong-coupling reduction. A similar mechanism is also expected in three dimensions. Since the interfacial area per unit volume still scales as $L^{-1}$ during dynamical scaling, the availability of compatible pairs again decreases as $L^{-2}$, suggesting the same competition between cooperative and local interface motion, although the morphology and connectivity of domain-wall surfaces may introduce additional kinetic effects.

More broadly, related anomalous coarsening may arise whenever slow classical variables are coupled to microscopic electrons and the motion of defects requires a redistribution of electronic charge. This constraint is particularly consequential for extended domain walls and interfaces, whose local displacement can require a compensating electronic rearrangement at a distant part of the defect network. Candidate settings include structural, charge, and orbital ordering transitions in which lattice distortions or other classical collective coordinates evolve in the presence of a conserved electronic density. In such systems, the coarsening kinetics need not be determined solely by whether the coarse-grained order parameter is conserved; constraints inherited from the microscopic electronic degrees of freedom can generate a defect-density-dependent mobility and long crossovers between apparent growth laws. It would therefore be useful to examine directly the correlations between spatially separated wall motions and to determine how the crossover scale evolves with electron hopping, filling, interaction strength, and dimensionality.

\begin{acknowledgments}
This work was supported by the US Department of Energy Basic Energy Sciences under Contract No. DE-SC0020330.   The authors also acknowledge the support of Research Computing at the University of Virginia.
\end{acknowledgments}

\appendix

\section{Dimensionless units}
\label{app:units}

For the numerical simulations, it is convenient to express the quasiclassical Holstein model in dimensionless form. This removes redundant microscopic energy and length scales and makes clear which combinations of parameters control the dynamics. The natural scales follow directly from the competition between the local elastic energy and the electron--lattice coupling.

Consider the dimensional quasiclassical Hamiltonian
\begin{equation}
\mathcal{H}
=
\sum_i\left(
\frac{P_i^2}{2M}
+\frac{K}{2}Q_i^2
\right)
+\kappa\sum_{\langle ij\rangle}Q_iQ_j
-g\sum_i
\left(n_i-\frac{1}{2}\right)Q_i .
\label{eq:dimensional_hamiltonian}
\end{equation}
Balancing the onsite elastic force, $KQ_i$, against the electronic force of order $g$ identifies the characteristic displacement scale
\begin{equation}
Q_0=\frac{g}{K}.
\label{eq:Q0}
\end{equation}
The corresponding energy scale is
\begin{equation}
E_0=KQ_0^2=\frac{g^2}{K}.
\label{eq:E0}
\end{equation}
Physically, $E_0$ is the characteristic local lattice-relaxation or polaronic energy associated with the electron--lattice coupling. We use $Q_0$ and $E_0$ as the basic displacement and energy units throughout the simulations.

The natural frequency of the local oscillator is $\Omega_0=\sqrt{K/M}$, which defines the time scale
\begin{equation}
t_0=\Omega_0^{-1}=\sqrt{\frac{M}{K}}.
\label{eq:t0}
\end{equation}
The associated momentum scale is $P_0=M Q_0/t_0=g\sqrt{M/K}$. We therefore introduce the dimensionless variables
\begin{equation}
q_i=\frac{Q_i}{Q_0}=\frac{K}{g}Q_i,
\qquad
p_i=\frac{P_i}{P_0}
=\frac{P_i}{g\sqrt{M/K}},
\qquad
\tau=\frac{t}{t_0},
\label{eq:dimensionless_variables}
\end{equation}
together with the dimensionless intersite coupling
\begin{equation}
\lambda=\frac{\kappa}{K}.
\label{eq:lambda}
\end{equation}
Thus $\lambda$ measures the strength of the nearest-neighbor elastic coupling relative to the onsite stiffness. The value of $\lambda$, rather than $K$ and $\kappa$ separately, controls the relative energetic preference for staggered distortions in the dimensionless model.

Dividing Eq.~\eqref{eq:dimensional_hamiltonian} by $E_0=g^2/K$ gives
\begin{equation}
\frac{\mathcal{H}}{E_0}
=
\sum_i\frac{p_i^2}{2}
+\frac{1}{2}\sum_i q_i^2
+\lambda\sum_{\langle ij\rangle}q_iq_j
-\sum_i\left(n_i-\frac{1}{2}\right)q_i .
\label{eq:dimensionless_hamiltonian}
\end{equation}
In the following, we denote the dimensionless Hamiltonian $\mathcal{H}/E_0$ simply by $\mathcal{H}$ when no confusion can arise.

The electronic quantities are rescaled by the same energy unit. In particular, the local electronic level in the quasiclassical limit, $\epsilon_i=-gQ_i$, becomes
\begin{equation}
\varepsilon_i\equiv\frac{\epsilon_i}{E_0}=-q_i.
\label{eq:dimensionless_local_energy}
\end{equation}
Similarly, we define the dimensionless temperature and chemical potential by
\begin{equation}
T\rightarrow\frac{k_{\rm B}T}{E_0},
\qquad
\mu\rightarrow\frac{\mu}{E_0}.
\label{eq:dimensionless_temperature}
\end{equation}
With this convention, the Fermi--Dirac occupation retains the same form as in the dimensional problem,
\begin{equation}
n_i
=
\frac{1}
{\exp[(\varepsilon_i-\mu)/T]+1}
=
\frac{1}
{\exp[(-q_i-\mu)/T]+1},
\label{eq:dimensionless_fermi}
\end{equation}
where $\mu$ is adjusted at each time step to enforce half filling.

The dimensionless force associated with the coordinate $q_i$ follows from the negative gradient of the dimensionless energy,
\begin{equation}
f_i
\equiv
-\frac{\partial\mathcal{H}}{\partial q_i}
=
-q_i
-\lambda\sum_{j\in{\rm nn}(i)}q_j
+\left(n_i-\frac{1}{2}\right).
\label{eq:dimensionless_force}
\end{equation}
The three contributions have a transparent interpretation. The first is the onsite elastic restoring force, the second arises from the nearest-neighbor elastic coupling, and the final term is the electronic force associated with the local deviation from half filling. In the quasiclassical limit, the occupation $n_i$ is determined directly from Eq.~\eqref{eq:dimensionless_fermi}, so the electronic contribution is a local function of $q_i$ once the global chemical potential has been determined.

For completeness, the Langevin equation may also be written entirely in these units. Introducing the dimensionless damping coefficient
\begin{equation}
\tilde{\gamma}
=
\frac{\gamma}{\sqrt{MK}},
\label{eq:dimensionless_gamma}
\end{equation}
the equation of motion takes the form
\begin{equation}
\frac{d^2q_i}{d\tau^2}
=
f_i
-\tilde{\gamma}\frac{dq_i}{d\tau}
+\xi_i(\tau),
\label{eq:dimensionless_langevin}
\end{equation}
where the rescaled stochastic force satisfies
\begin{equation}
\left\langle
\xi_i(\tau)\xi_j(\tau')
\right\rangle
=
2\tilde{\gamma}T\,
\delta_{ij}\delta(\tau-\tau').
\label{eq:dimensionless_noise}
\end{equation}
Thus, after rescaling, the basic dynamics is controlled by a small number of dimensionless parameters, most importantly the elastic-coupling ratio $\lambda$, damping $\tilde{\gamma}$, and temperature $T$. The simulations discussed in the main text use these dimensionless units; in particular, quoted temperatures are measured in units of $g^2/K$ and times in units of $\sqrt{M/K}$.

\bibliography{ref}
\end{document}